\documentclass[runningheads]{llncs}
\usepackage[T1]{fontenc}
\usepackage{graphicx}
\usepackage{booktabs, tabularx}
\usepackage{array}

\usepackage{algorithmicx}
\usepackage{mathtools}
\usepackage{subcaption}
\usepackage{amsmath,amsfonts,stackrel}
\usepackage[utf8]{inputenc} % allow utf-8 input
\usepackage[T1]{fontenc}    % use 8-bit T1 fonts
\usepackage{hyperref}       % hyperlinks
\usepackage{url}            % simple URL typesetting
\usepackage{booktabs}       % professional-quality tables
\usepackage{amsfonts}       % blackboard math symbols
\usepackage{adjustbox}
\usepackage{nicefrac}       % compact symbols for 1/2, etc.
\usepackage{microtype}      % micro typography
\usepackage{lipsum}
\usepackage{fancyhdr}       % header
\usepackage{tabularx}
\graphicspath{{media/}}     % organize your images and other figures under media/ folder
\usepackage{babel}
\usepackage{etoolbox}
\usepackage{xcolor}
\usepackage{amssymb}
\usepackage{algorithm}
\usepackage{algpseudocode}
\usepackage[flushleft]{threeparttable}
\usepackage{mathtools}

\usepackage{appendix}
\usepackage{xcolor}
\usepackage{booktabs}
\usepackage{subcaption}
\usepackage{multirow}

\usepackage{subcaption}

\newcommand{\R}{\mathbb{R}}
\newcommand{\prob}{\mathbb{P}}

\newcommand{\E}{\mathbb{E}}

\newcommand{\pto}{\xlongrightarrow{p}}
\newcommand{\dto}{\xlongrightarrow{d}}
\usepackage{float}

\DeclareMathOperator{\Var}{Var}

\newcolumntype{C}{>{\centering\arraybackslash}X}

\begin{document}
\title{Differentially Private Confidence Intervals for Relative Risk and Odds Ratio}
%
%\titlerunning{Differential privacy and ratio statistics}
% If the paper title is too long for the running head, you can set
% an abbreviated paper title here
%
\author{Tomer Shoham \and
Katrina Ligett}
\authorrunning{T. Shoham and K. Ligett}
\institute{
School of Computer Science and Engineering, The Hebrew University of Jerusalem\
\email{{Tomer.Shoham,Katrina.Ligett}@mail.huji.ac.il}
}

\maketitle              % typeset the header of the contribution
\begin{abstract}
Ratio statistics—such as relative risk and odds ratio—are central tools for comparing outcome probabilities across populations, particularly in clinical and epidemiological studies. However, when the underlying counts are released under differential privacy, the resulting ratio estimators can exhibit nontrivial bias and variance, and principled methods for statistical inference in this setting remain limited. 
In this work, we study estimation and inference for ratio statistics computed from counts perturbed by additive Laplace or Gaussian noise. We show that the natural plug-in estimators based on privatized counts are consistent, and we derive confidence intervals that explicitly account for both sampling variability and privacy-induced noise, making our approach appealing for moderate sample sizes. Our analysis provides transparent variance decompositions that clarify when privacy meaningfully affects inference. Empirical results on simulated and real datasets demonstrate that, for moderate privacy levels and sample sizes, the proposed confidence intervals closely match their non-private counterparts, while properly reflecting additional uncertainty in small-sample regimes. Our results provide both theoretical guarantees and practical guidance for conducting statistical inference with ratio statistics under differential privacy.

\end{abstract}

\section{Introduction}

In many practical settings, analysts do not design the privacy mechanism themselves. Instead, they receive counts that have already been perturbed to satisfy differential privacy (DP), for example, by a statistical agency, hospital, or data curator, and must carry out standard downstream analyses using these noisy releases. A common example is the analysis of $2\times2$ contingency tables, where one wishes to compare event probabilities across two groups using ratio statistics such as the \emph{relative risk} (RR) and the \emph{odds ratio} (OR). These quantities are central in clinical trials, epidemiology, and public health, and they also arise broadly whenever event rates are compared across populations.

This raises a basic practical question: if the underlying counts have already been privatized, can one still form statistically meaningful estimates and confidence intervals (CIs) for RR and OR by plugging in the noisy counts? Since ratio statistics can be unstable when denominators are small, and since additive DP noise further perturbs the relevant counts, it is not a priori clear when such post-processed estimators remain reliable.

This paper is meant as a practical guide to that question. We study estimation and inference for RR and OR when the underlying counts are released under additive DP noise. We prove consistency of the private plug-in estimators (Lemma \ref{lem:consist_est}), derive asymptotically valid CIs under Gaussian perturbation (Theorems \ref{thm:ci_rr_gauss}--\ref{thm:ci_rr_direct}), and show how the same formulas extend naturally to Laplace noise by variance substitution. Our variance expressions separate sampling- and privacy-induced variability (Table \ref{table:var_decom}), yielding a simple, practical rule: privacy has limited impact when effective cell counts are large, and becomes consequential mainly in sparse regimes. Our main message is simple: privacy has the largest effect precisely where inference was already difficult to begin with. Simulations and real-data examples (Section \ref{sec:num_study}) support this picture, showing that, for moderate privacy levels and sample sizes, the private CIs are typically close to their non-private counterparts. 

\textbf{Related work.} \ Our perspective is closely connected to a growing literature on inference from privatized low-dimensional summaries. A substantial body of work studies DP methods for frequency tables and related contingency-table problems~\cite{barak2007privacy,rinott2018confidentiality,vu2009differential,gaboardi2016differentially,awan2018differentially}. More recently, the deployment of DP in official statistics, especially by the U.S.~Census Bureau for the 2020 Census, has sharpened attention on the needs of analysts who must consume already-noised releases and understand how privacy propagates into downstream estimands~\cite{mervis2019can,mueller20222020,santos2020differential}. Our work fits squarely into this consumer-of-DP-data viewpoint.

Most closely related to our work is a paper by Lin et al.~\cite{lin2024differentially}, wherein they study the problem of constructing CIs around proportion estimates when using stratified random sampling techniques under DP constraints. In situations where the denominator of their proportion estimate (the size of the relevant sample) is private information, they consider adding Gaussian noise for DP to both the numerator and the denominator. As a result, they end up with a mathematical object---a ratio of two Gaussian random variables---that is also central to our analysis. They provide a detailed analysis of this ratio and derive an asymptotically valid CI.
%To arrive at this result, they leverage results of D{\'\i}az-Franc{\'e}s and Rubio~\cite{diaz2013existence} on the goodness of approximation of the ratio of two Gaussian random variables, in particular a Berry-Esseen-like theorem bounding the total variation distance between the ratio of two Gaussians and a single Gaussian distribution. 
In Section~\ref{sec:CI}, we use proof techniques similar to those of \cite{lin2024differentially} to obtain another DP-CI for the RR (Theorem \ref{thm:ci_rr_direct}).

A longer version of this work appears on arXiv~\cite{shoham2025differentially}. The full version includes additional mechanisms and extensions, including direct perturbation of the ratio, transformed estimators, local-sensitivity-based approaches, and a more detailed treatment of bias and finite-sample behavior.

\section{Preliminaries}\label{sec:2}
Unless stated otherwise, $\prob$ denotes probability with respect to the joint distribution over all sources of randomness. We use $\pto$ for convergence in probability, and $\dto$ for convergence in distribution. Throughout the paper, non-private estimators are denoted by $\widehat{\cdot}$ and their differentially private counterparts by $\widetilde{\cdot}$.

%Let $X_n$ and $X$ be random variables defined on a common probability space and taking values in a metric space. We write $X_n \pto X$ (convergence in probability) if for every $\xi>0$, $\lim_{n\to\infty} \prob(|X_n - X| > \xi) = 0$. We write $X_n \dto X$ (convergence in distribution) if $\lim_{n\to\infty} F_{X_n}(x) = F_X(x)$ for every continuity point $x$ of $F_X$, where $F_{X_n}$ and $F_X$ denote the cumulative distribution functions of $X_n$ and $X$, respectively. We use standard stochastic order notation. For a sequence of random variables $X_n$ and positive deterministic scalars $a_n$, we write $X_n = O_p(a_n)$ if $X_n/a_n$ is bounded in probability, and $X_n = o_p(a_n)$ if $X_n/a_n \pto 0$.

\subsection{The Relative Risk and Odds Ratio Statistics}

The relative risk (RR) and odds ratio (OR) are fundamental measures for comparing outcome probabilities between two groups. The RR quantifies the multiplicative change in the probability of an event, while the OR compares the corresponding odds. When event probabilities are small, the two measures are numerically similar, but they may differ substantially when probabilities are moderate or large. The RR is often preferred for interpretability in clinical settings, whereas the OR arises naturally in logistic modeling and case–control studies; see Agresti~\cite{agresti2012categorical} for background.

We suppose that $X$ and $Y$ are independent binomial random variables, $X \sim \mathrm{Bin}(n_x,p_x)$ and $Y \sim \mathrm{Bin}(n_y,p_y)$, where $p_x,p_y\in(0,1)$ and $n_x,n_y\in\mathbb{N}$ with $n_x+n_y=n$. The sample sizes $n_x$ and $n_y$ are assumed known. Our goal is to estimate either $RR \coloneqq p_x/p_y$ or $OR \coloneqq \frac{p_x/(1-p_x)}{p_y/(1-p_y)}$.

Let $\widehat p_x \coloneqq X/n_x$ and $\widehat p_y \coloneqq Y/n_y$ denote the maximum likelihood estimators of $p_x$ and $p_y$. The corresponding plug-in estimators are 
$$\widehat{RR} \coloneqq \widehat p_x/\widehat p_y \qquad \text{and} \qquad \widehat{OR} \coloneqq \frac{\widehat p_x/(1-\widehat p_x)}{\widehat p_y/(1-\widehat p_y)}.$$ These estimators are biased on finite samples but consistent for their respective population parameters; see, for example, van~der~Vaart~\cite{van2000asymptotic}. 

To quantify uncertainty, one typically constructs CIs.
\begin{definition}[Confidence Interval]\label{def:CI}
Let $F_\theta$ be a distribution on $\mathbb{R}$ and let $X=(X_1,\dots,X_n)$ be an i.i.d.\ sample from $F_\theta$. A $(1-\alpha)$ confidence interval for $\theta$ is a data-dependent interval $(u(X),v(X))$ satisfying 
$$\lim_{n\to\infty} \prob(u(X)\le \theta \le v(X)) = 1-\alpha,$$ 
where probability is taken with respect to both the sampling randomness and any additional randomness in $u(\cdot)$ and $v(\cdot)$.
\end{definition}

Classical large-sample CIs for RR and OR are typically derived via normal approximations and the delta method; see Agresti~\cite[Ch.~2]{agresti2012categorical}.

\subsection{Differential privacy}

We briefly review the notions from differential privacy that are used throughout the paper; see Dwork and Roth~\cite{dwork2014algorithmic} for a comprehensive treatment.

Let $\cal X$ be a data domain. An unordered dataset of size $n$ is represented as a tuple $D=(D_1,\dots,D_n)\in\mathcal X^n$, where $n$ is assumed public. Two datasets $D,D'\in{\cal X}^n$ are called \emph{neighbors}, written $D\sim D'$, if they differ in exactly one entry. DP requires that the output distribution of a randomized algorithm not change too much when applied to neighboring datasets.

\begin{definition}[Differential privacy]
A randomized algorithm ${\cal M}:{\cal X}^n\to{\cal Y}$ is $(\varepsilon,\delta)$-differentially private if for all neighboring datasets $D\sim D'$ and all measurable sets $S\subseteq{\cal Y}$, $
\prob\!\big({\cal M}(D)\in S\big)
\le
e^\varepsilon\,\prob\!\big({\cal M}(D')\in S\big)+\delta$.
\end{definition}

Smaller values of $\varepsilon$ and $\delta$ correspond to stronger privacy guarantees. Two basic properties of DP are especially relevant in practice: privacy guarantees are preserved under arbitrary post-processing of the released output, and they degrade in a controlled way under composition of multiple private mechanisms; see Proposition~2.1 and Theorem~3.14 in~\cite{dwork2014algorithmic}. The post-processing property is particularly important for our purposes, since throughout the paper we analyze ratio statistics constructed from already-privatized counts.

A standard way to ensure differential privacy for numerical queries is to add
random noise calibrated to the query's sensitivity. For
$f:{\cal X}^n\to\mathbb R^k$, define the $\ell_1$- and $\ell_2$-sensitivities by
$
\Delta_{f,1} \coloneqq \max_{D\sim D'} \|f(D)-f(D')\|_1$, $\Delta_{f,2} \coloneqq \max_{D\sim D'} \|f(D)-f(D')\|_2.
$
The following two classical additive-noise mechanisms will be used throughout.

\begin{theorem}[Laplace and Gaussian mechanisms]\label{thm:additive_noise_dp}
Let $f:{\cal X}^n\to\mathbb R^k$. Consider mechanisms of the form
$
{\cal M}_f(D)=f(D)+W,
$
where $W\in\mathbb R^k$ is a random noise vector independent of the data.
Then the following guarantees hold.
\begin{itemize}
    \item \textbf{Laplace mechanism:} If
    $
    W=(W_1,\ldots,W_k)
    $
    has independent coordinates with
    $
    W_j\sim \mathrm{Lap}(\Delta_{f,1}/\varepsilon)$, $j=1,\ldots,k$,
    then ${\cal M}_f$ is $(\varepsilon,0)$-DP
    \cite[Thm.~3.6]{dwork2014algorithmic}.

    \item \textbf{Gaussian mechanism:} If
    $
    W\sim \mathcal N(0,\sigma^2 I_k),
    $
    then ${\cal M}_f$ is $(\varepsilon,\delta)$-DP if and only if
    $
    \Phi\!\left(\frac{\Delta_{f,2}}{2\sigma}
    -\frac{\varepsilon\sigma}{\Delta_{f,2}}\right)
    -
    e^\varepsilon
    \Phi\!\left(-\frac{\Delta_{f,2}}{2\sigma}
    -\frac{\varepsilon\sigma}{\Delta_{f,2}}\right)
    \le \delta,
    $
    where $\Phi$ is the standard normal cumulative distribution function
    \cite[Thm.~8]{balle2018improving}.
\end{itemize}
\end{theorem}

\section{Theoretical results}\label{sec:CI}

In this section, we characterize the joint impact of sampling variability and privacy-induced noise on the RR and OR. While classical CIs remain asymptotically valid when privacy perturbations are of a lower order than sampling error, they become insufficient in finite-sample regimes where privacy noise significantly alters the error distribution. We derive CIs for both RR and OR under Gaussian noise and establish their asymptotic validity. Finally, we demonstrate how this construction extends to the Laplace mechanism, providing a general framework for inference with privatized count data.

\subsection{Setting}\label{subsec:setting}

We consider independent binomial random variables 
$X \sim \mathrm{Bin}(n_x,p_x)$ and 
$Y \sim \mathrm{Bin}(n_y,p_y)$ with $p_x,p_y\in(0,1)$ and known sample sizes $n_x,n_y>1$. 
Privatized counts are released as 
$\widetilde X = X + N_x$ and 
$\widetilde Y = Y + N_y$, 
where $(N_x,N_y)$ are independent of $(X,Y)$ and of each other.

Our analysis uses replacement adjacency: the sample sizes $n_x$ and $n_y$ are treated as public, and neighboring databases differ in the event of one individual. If $n_x,n_y$ are not public, one could instead estimate them privately, using part of the privacy budget. The RR and OR estimators could then be computed by post-processing these noisy counts. The resulting CIs would need to account for the additional uncertainty, but this extension does not conceptually change our results and is therefore omitted due to space limitations.

In this section, whenever we write $\pto$, $\dto$, $O(\cdot)$, or $o(\cdot)$, the asymptotic regime is understood to be
$
n_x\to\infty$, $n_y\to\infty$, $\frac{n_x}{n_x+n_y}\to\lambda\in(0,1)
$.

We consider either (i) Laplace noise $N_x \sim \mathrm{Lap}(b_x)$, $N_y \sim \mathrm{Lap}(b_y)$ or (ii) Gaussian noise $N_x \sim \mathcal N(0,\sigma_x^2)$, $N_y \sim \mathcal N(0,\sigma_y^2)$, with parameters calibrated according to Theorem~\ref{thm:additive_noise_dp}.  
We consider $\Var(N_x)=o(n_x^2)$ and $\Var(N_y)=o(n_y^2)$; equivalently $\mathrm{sd}(N_x)=o(n_x)$ and $\mathrm{sd}(N_y)=o(n_y)$. This allows privacy perturbations that may exceed the usual sampling fluctuations of order $\sqrt{n_x}$ and $\sqrt{n_y}$, while still guaranteeing consistency. For example, the assumption $\Var(N)=o(n^2)$ implies that if we were to use the Laplace mechanism (Theorem~\ref{thm:additive_noise_dp}), we assume that $\varepsilon = \omega(n^{-1})$.

Denote $\widetilde p_x=\widetilde X/n_x$ and $\widetilde p_y=\widetilde Y/n_y$. 
To ensure ratios are well defined, all statistics are computed using projected counts 
$\widetilde X_p=\min(\max(1,\widetilde X),n_x-1)$ and 
$\widetilde Y_p=\min(\max(1,\widetilde Y),n_y-1)$\footnote{While we use a clipping constant of 1 for simplicity, our empirical tests (omitted for brevity) show that for $n > 50$, the coverage is remarkably robust to the choice of $\text{clip} \in [0.1, 1]$}.
Under $p_x,p_y\in[\eta,1-\eta]$ for some fixed $\eta>0$, together with $\Var(N_x)=o(n_x^2)$ and $\Var(N_y)=o(n_y^2)$, we have $\prob(\widetilde X\notin[1,n_x-1])+\prob(\widetilde Y\notin[1,n_y-1])=o(1)$.
Thus, clipping affects the statistics with vanishing probability, and we omit the subscript $p$ for brevity.

With these definitions, we consider the private plug-in estimators 
\begin{equation}\label{eq:priv_RR_OR}
\widetilde{RR} \coloneqq \widetilde p_x/\widetilde p_y \qquad \widetilde{OR} \coloneqq \frac{\widetilde p_x/(1-\widetilde p_x)}{\widetilde p_y/(1-\widetilde p_y)}.
\end{equation}

\begin{lemma}[Consistency of the perturbed RR and OR]\label{lem:consist_est}
Under the standing assumptions of Subsection~\ref{subsec:setting}, $\widetilde{RR}\pto RR$ and $\widetilde{OR}\pto OR$.
\end{lemma}

\begin{proof}
We write 
$\widetilde p_x - p_x = (X-n_xp_x)/n_x + N_x/n_x$. 
By the weak law of large numbers, $(X-n_xp_x)/n_x \pto 0$. 
Since $\Var(N_x)=o(n_x^2)$, Chebyshev's inequality implies $N_x/n_x \pto 0$. 
Hence $\widetilde p_x \pto p_x$, and similarly $\widetilde p_y \pto p_y$.

The mappings $g(u,v)=u/v$ and 
$h(u,v)=\frac{u/(1-u)}{v/(1-v)}$ 
are continuous at $(p_x,p_y)$ when $p_x,p_y\in(0,1)$. 
The result follows from the continuous mapping theorem.
\end{proof}

\subsection{Confidence Intervals}\label{subsec:CI}

We derive asymptotically valid CIs under additive Gaussian noise. 
All results are stated under the standing assumptions of Subsection~\ref{subsec:setting}. 

For a real-valued statistic $T$ and standard error estimate $\widehat{\mathrm{se}}(T)$, define the Wald-type interval
$
\mathcal C_{1-\alpha}(T,\widehat{\mathrm{se}})
\coloneqq
\left[T \pm z_{1-\alpha/2}\widehat{\mathrm{se}}\right],
$
where $z_{1-\alpha/2}$ is the $(1-\alpha/2)$ quantile of a standard normal.

With this in mind, we now state our main theorems for valid CIs under Gaussian noise addition.

\begin{theorem}\label{thm:ci_rr_gauss}
Under the standing assumptions of Subsection~\ref{subsec:setting}, define
$$
\widetilde \ell_{RR}
=
\log\!\left(\frac{\widetilde X/n_x}{\widetilde Y/n_y}\right),
\qquad
\ell^*_{RR}
=
\log\!\left(\frac{p_x}{p_y}\right),
$$
$$
\widehat{\mathrm{se}}_{RR}^2
=
\Big(\frac{1}{\widetilde X}-\frac{1}{n_x}\Big)
+
\Big(\frac{1}{\widetilde Y}-\frac{1}{n_y}\Big)
+
\sigma_x^2\frac{1}{\widetilde X^2}
+
\sigma_y^2\frac{1}{\widetilde Y^2}.
$$
Then the interval
$
\mathcal C_{1-\alpha}^{\log RR}
\coloneqq
\mathcal C_{1-\alpha}(\widetilde \ell_{RR},\widehat{\mathrm{se}}_{RR})
$
satisfies
$
\prob(\ell^*_{RR}\in\mathcal C_{1-\alpha}^{\log RR})\to1-\alpha.
$
Consequently,
$
\mathcal C_{1-\alpha}^{RR}
\coloneqq
\exp\!\big(\mathcal C_{1-\alpha}^{\log RR}\big)
$
is an asymptotic $(1-\alpha)$ CI for $RR$.
\end{theorem}
An analogous result can be derived for the OR statistic.
\begin{theorem}\label{thm:ci_or_gauss}
Under the standing assumptions of Subsection~\ref{subsec:setting}, define
$$
\widetilde \ell_{OR}
=
\log\!\left(\frac{\widetilde X}{n_x-\widetilde X}\right)
-
\log\!\left(\frac{\widetilde Y}{n_y-\widetilde Y}\right),
\quad
\ell^*_{OR}
=
\log\!\left(\frac{p_x}{1-p_x}\right)
-
\log\!\left(\frac{p_y}{1-p_y}\right),
$$
$$
\widehat{\mathrm{se}}_{OR}^2
=
\Big(\frac{1}{\widetilde X}+\frac{1}{n_x-\widetilde X}\Big)
+
\Big(\frac{1}{\widetilde Y}+\frac{1}{n_y-\widetilde Y}\Big)
+
\sigma_x^2\Big(\frac{1}{\widetilde X}+\frac{1}{n_x-\widetilde X}\Big)^2
+
\sigma_y^2\Big(\frac{1}{\widetilde Y}+\frac{1}{n_y-\widetilde Y}\Big)^2.
$$
Then the interval
$
\mathcal C_{1-\alpha}^{\log OR}
\coloneqq
\mathcal C_{1-\alpha}(\widetilde \ell_{OR},\widehat{\mathrm{se}}_{OR})
$
satisfies
$
\prob(\ell^*_{OR}\in\mathcal C_{1-\alpha}^{\log OR})\to1-\alpha.
$
Consequently,
$
\mathcal C_{1-\alpha}^{OR}
\coloneqq
\exp\!\big(\mathcal C_{1-\alpha}^{\log OR}\big)
$
is an asymptotic $(1-\alpha)$ CI for $OR$.
\end{theorem}

The proofs of Theorems~\ref{thm:ci_rr_gauss} and~\ref{thm:ci_or_gauss} are given in the Appendix. At a high level, the argument has three steps: first, clipping the noisy counts is asymptotically negligible under $p_x,p_y\in[\eta,1-\eta]$ and $\sigma^2_x=o(n^2_x)$, $\sigma^2_y=o(n^2_y)$; second, the privatized counts are asymptotically normal (binomial + independent Gaussian noise); and third, applying the delta method to the log-RR and log-OR yields asymptotic normality with the stated variances. The proposed standard-error formulas are then shown to be consistent plug-in estimators of these asymptotic variances, and Slutsky's theorem gives the validity of the resulting CIs.

\subsubsection{A Direct Normal Confidence Interval for the Relative Risk}

In addition to the log-scale construction above, we also establish a CI based on a direct normal approximation of the ratio itself. While related in spirit to the approach of \cite{lin2024differentially}, our result is tailored to the two-sample setting with known group sizes. For completeness, we state the result below; a full proof appears as Theorem~4.2 in our extended manuscript~\cite{shoham2025differentially}.

\begin{theorem}[Direct normal CI for $RR$]\label{thm:ci_rr_direct}
Under the standing assumptions of Subsection~\ref{subsec:setting}, let
$
\widetilde{RR}=\frac{\widetilde X/n_x}{\widetilde Y/n_y},
$
and define
$$
\widehat{\mathrm{se}}_{RR}^2
=
\widetilde{RR}^{\,2}
\left(
\frac{1}{\widetilde X}-\frac{1}{n_x}
+
\frac{1}{\widetilde Y}-\frac{1}{n_y}
+
\frac{\sigma_x^2}{\widetilde X^2}
+
\frac{\sigma_y^2}{\widetilde Y^2}
\right).
$$
Then the interval
$
\mathcal C_{1-\alpha}^{RR}
\coloneqq
\mathcal C_{1-\alpha}(\widetilde{RR},\widehat{\mathrm{se}}_{RR})
$
satisfies
$
\prob(RR\in\mathcal C_{1-\alpha}^{RR})\to1-\alpha.
$
\end{theorem}

\textbf{Laplace perturbations.}
The preceding theorems are stated for Gaussian perturbations, for which the privatized counts remain asymptotically normal after variance adjustment. For Laplace perturbations, the same plug-in variance formulas can be used as a finite-sample approximation by replacing $\sigma_x^2,\sigma_y^2$ with $2b_x^2,2b_y^2$. Formally, if $b_x^2=o(n_x)$ and $b_y^2=o(n_y)$, so that the privacy perturbation is lower order than the sampling error, then the Laplace noise is asymptotically negligible and the resulting intervals remain asymptotically valid. When the Laplace noise is not lower order than the sampling error, asymptotic normality no longer follows from negligibility of the privacy perturbation. In this case, our interval is best understood as a moment-matched Gaussian approximation, obtained by replacing the Laplace noise with Gaussian noise of the same variance.

\subsection{When does privacy materially affect inference?}\label{subsec:connection}

A central concern for analysts using DP statistics is whether privacy-induced perturbation can severely reduce statistical power. Our variance formulas (Table \ref{table:var_decom}) show that, in the settings we study, this concern is often less severe than it may first appear.

Table~\ref{table:var_decom} decomposes the asymptotic variance into a \emph{sampling} component and a \emph{perturbation} component, for simplicity setting $\sigma_x=\sigma_y\coloneqq \sigma$. The key structural feature is that the sampling terms scale like $1/\widetilde X$ and $1/\widetilde Y$, whereas the perturbation terms scale like $\sigma^2/\widetilde X^2$ and $\sigma^2/\widetilde Y^2$. Thus, for fixed $(\varepsilon,\delta)$, the sampling standard error decreases at rate $O(n^{-1/2})$, while the additional error due to privacy decreases faster, at rate $O(n^{-1})$.

This gap in rates has an important consequence. In high-power regimes, where the counts are large, privacy adds only a lower-order term to the overall uncertainty. Conversely, when the privacy term is non-negligible, the sampling variance is already large, meaning that the corresponding non-private statistic is itself unstable. Thus, a large privacy-induced contribution to the interval width should not be interpreted as privacy alone creating the difficulty; rather, it indicates a regime in which the underlying non-private inference is already fragile. Indeed, taking the ratio between the DP CI width (sampling plus perturbation error) and the non-private width (sampling error only), we obtain
\begin{equation*}\label{eq:RR_width_ratio}
\frac{\mathrm{Width}_{DP}^{\log RR}}{\mathrm{Width}_{NP}^{\log RR}}
\approx
\frac{\mathrm{Width}_{DP}^{RR}}{\mathrm{Width}_{NP}^{RR}}
\approx
\sqrt{1+O\!\left(\frac{\sigma^2}{\min\{n_xp_x,\;n_yp_y\}}\right)}.
\end{equation*}
Similarly, for the odds ratio,
\begin{equation*}\label{eq:OR_width_ratio}
\frac{\mathrm{Width}_{DP}^{\log OR}}{\mathrm{Width}_{NP}^{\log OR}}
\approx
\sqrt{1+O\!\left(\frac{\sigma^2}{\min\{n_xp_x(1-p_x),\;n_yp_y(1-p_y)\}}\right)}.
\end{equation*}

%Equations~\eqref{eq:RR_width_ratio} and~\eqref{eq:OR_width_ratio} quantify the inflation in expected CI width due to privacy.

\newpage
\textbf{Estimating privacy-induced inflation.}
The same variance decomposition can also be used as a simple diagnostic after the privatized table has been released. Since both the sampling component and the perturbation component in Table~\ref{table:var_decom} are functions of the released noisy counts and the known noise variance, an analyst can estimate the relative contribution of privacy noise without needing to incur any additional privacy cost. In particular, one can compare the perturbation component to the sampling component. For Laplace perturbations, the same diagnostic is obtained by replacing the Gaussian variance by the Laplace variance. This provides a practical way to assess, from the released table itself, whether the privacy mechanism materially changes the inferential conclusion.

\begin{table}[]
\centering
\scriptsize
\setlength{\tabcolsep}{2pt}
\renewcommand{\arraystretch}{1.25}
\begin{tabularx}{\linewidth}{@{}l|c|c@{}}
\toprule
Statistic & Sampling error & Perturbation error \\
\midrule

Log RR
&
$\displaystyle 
\frac{1}{\widetilde X}-\frac{1}{n_x}
+
\frac{1}{\widetilde Y}-\frac{1}{n_y}
$
&
$\displaystyle 
\sigma^2 \left( \frac{1}{\widetilde X^2}
+
\frac{1}{\widetilde Y^2} \right)
$
\\
\addlinespace[0.6em]
RR
&
$\displaystyle 
\left(\frac{\widetilde X/n_x}{\widetilde Y/n_y}\right)^2
\left(
\frac{1}{\widetilde X}-\frac{1}{n_x}
+
\frac{1}{\widetilde Y}-\frac{1}{n_y}
\right)
$
&
$\displaystyle 
\sigma^2 \left(\frac{\widetilde X/n_x}{\widetilde Y/n_y}\right)^2
\left(
\frac{1}{\widetilde X^2}
+
\frac{1}{\widetilde Y^2}
\right)
$
\\
\addlinespace[0.6em]
Log OR
&
$\displaystyle 
\frac{1}{\widetilde X}+\frac{1}{n_x-\widetilde X}
+
\frac{1}{\widetilde Y}+\frac{1}{n_y-\widetilde Y}
$
&
$\displaystyle 
\sigma^2 \left[
\left(
\frac{1}{\widetilde X}+\frac{1}{n_x-\widetilde X}
\right)^2
+
\left(
\frac{1}{\widetilde Y}+\frac{1}{n_y-\widetilde Y}
\right)^2 \right]
$
\\
\bottomrule
\end{tabularx}
\caption{Decomposition of the asymptotic variances in Theorems~\ref{thm:ci_rr_gauss}--\ref{thm:ci_rr_direct} into sampling and perturbation components.}
\label{table:var_decom}
\end{table}

\vspace{-1cm}
\section{Empirical evaluation}\label{sec:num_study}

We empirically study the bias of the private estimators, as well as the width and coverage of the proposed CIs, on both simulated and real data. Throughout the simulation study we set $n_x=n_y$ for simplicity. We also ran additional experiments with unequal group sizes and observed qualitatively similar behavior; these results are omitted for brevity.
\vspace{-0.3cm}
\subsubsection*{Bias.}
In general, it is known that the RR and OR non-private estimators are biased. We examine how additive noise further affects this bias. We fix $n_x=n_y=150$ and consider three parameter configurations, representing decreasing, null, and increasing signal regimes, 
$(p_x,p_y)\in\{(1/3,2/3),(1/2,1/2),(2/3,1/3)\}$. For each configuration and each privacy level $\varepsilon\in[0.5,2]$, we run $N=10^5$ independent repetitions. In each repetition, we generate fresh data $X\sim\mathrm{Bin}(n_x,p_x)$ and $Y\sim\mathrm{Bin}(n_y,p_y)$, and then independently perturb the counts using either Gaussian or Laplace noise. Gaussian noise is calibrated using the analytic Gaussian mechanism to satisfy $(\varepsilon,\delta)$-DP with $\delta=10^{-5}$, while Laplace noise uses scale $b=1/\varepsilon$ (Theorem~\ref{thm:additive_noise_dp}). We then compute the privatized plug-in estimator for RR or OR from the perturbed counts, with projection to $[1,n-1]$, and estimate its expectation by averaging across repetitions. For reference, we also report the mean of the corresponding non-private plug-in estimator and the true parameter value.

\begin{figure}[t]
\centering
\includegraphics[width=\textwidth,height=0.72\textheight,keepaspectratio]{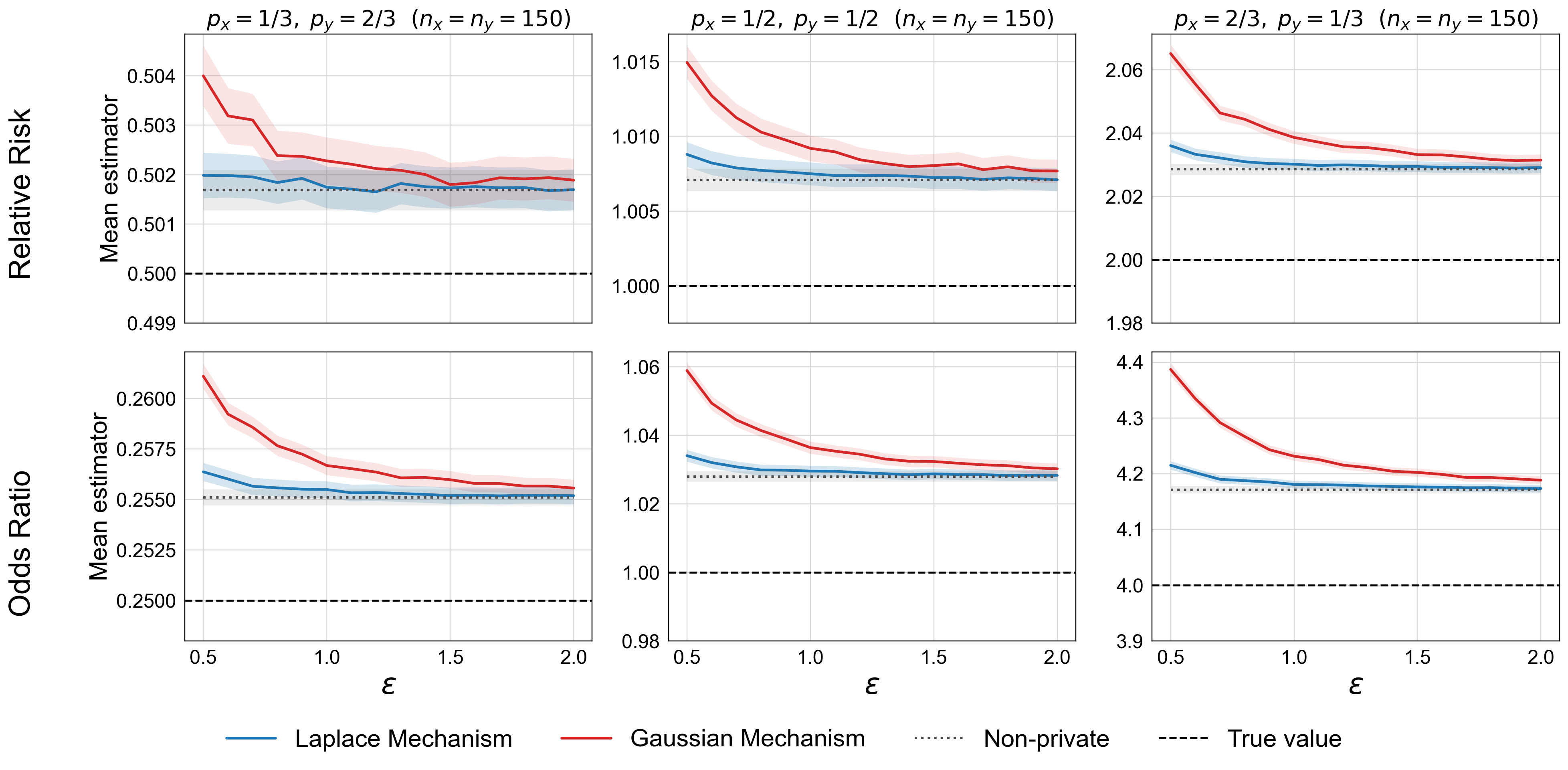}
\caption{Empirical mean of the privatized plug-in estimator for RR and OR \eqref{eq:priv_RR_OR} under Laplace and Gaussian noise addition. The dashed horizontal line indicates the true parameter value. Top row: RR. Bottom row: OR.}
\label{fig:RR_OR_mean_vs_true}
\end{figure}

Figure~\ref{fig:RR_OR_mean_vs_true} depicts the bias of the private and non-private estimators. The private estimators exhibit a small upward bias relative to their non-private counterparts, particularly in the strong-privacy regime (small $\varepsilon$). As expected, the bias decreases as $\varepsilon$ increases. Across all three parameter settings, the private estimators remain close to the non-private ones, and the discrepancy becomes negligible for moderate privacy levels. Notably, under Laplace noise addition, the bias induced by perturbation is of a lower order than the inherent sampling bias. For example, when $\varepsilon=1.25$, the bias due to privacy (the gap between the dotted line and the blue line) is negligible, while the bias due to sampling is clearly manifested (the gap between the dotted line and the dashed line).

\begin{figure}[t]
\centering

\begin{subfigure}{\textwidth}
\centering
\includegraphics[width=\textwidth,keepaspectratio]{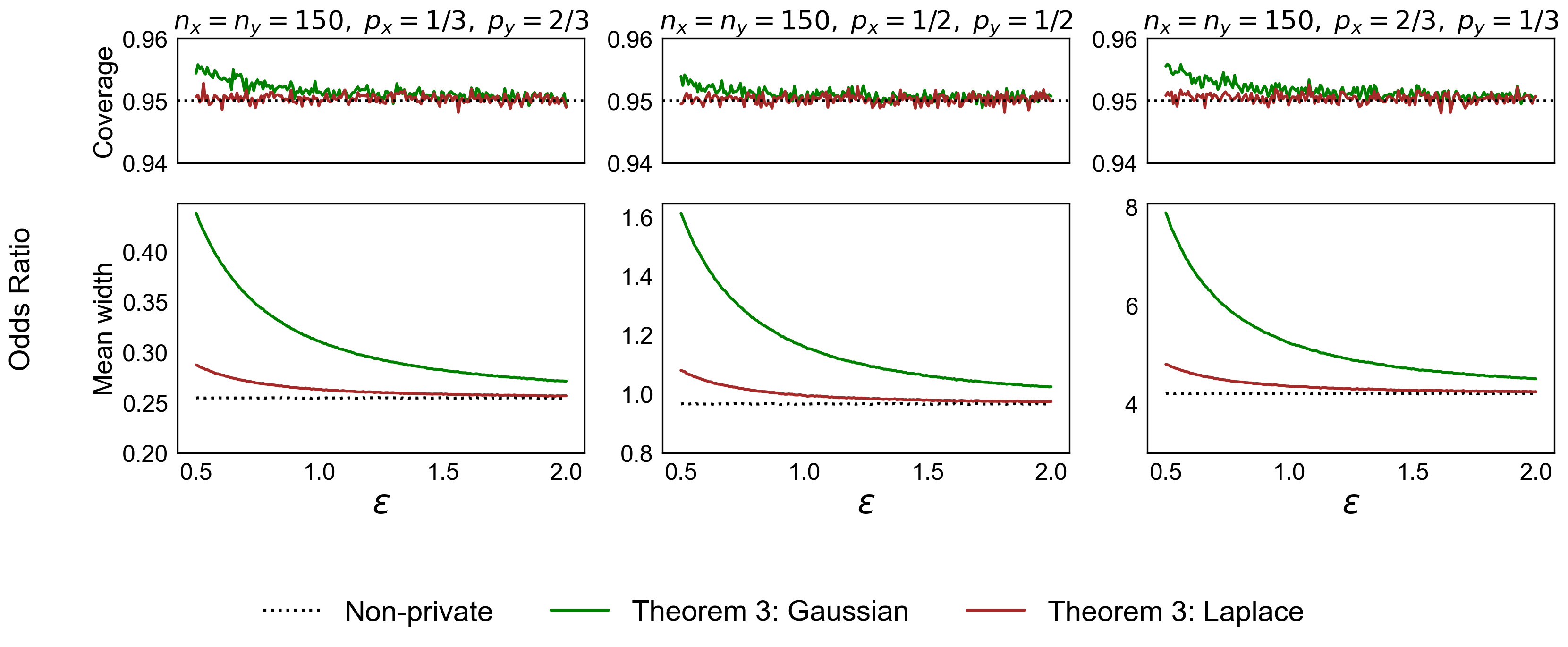}
\caption{\textbf{Odds Ratio}: Mean width and coverage for CI construction. The dashed horizontal line marks the nominal coverage $1-\alpha$.}
\label{fig:CI_width_coverage_OR}
\end{subfigure}

\vspace{1em}

\begin{subfigure}{\textwidth}
\centering
\includegraphics[width=\textwidth,keepaspectratio]{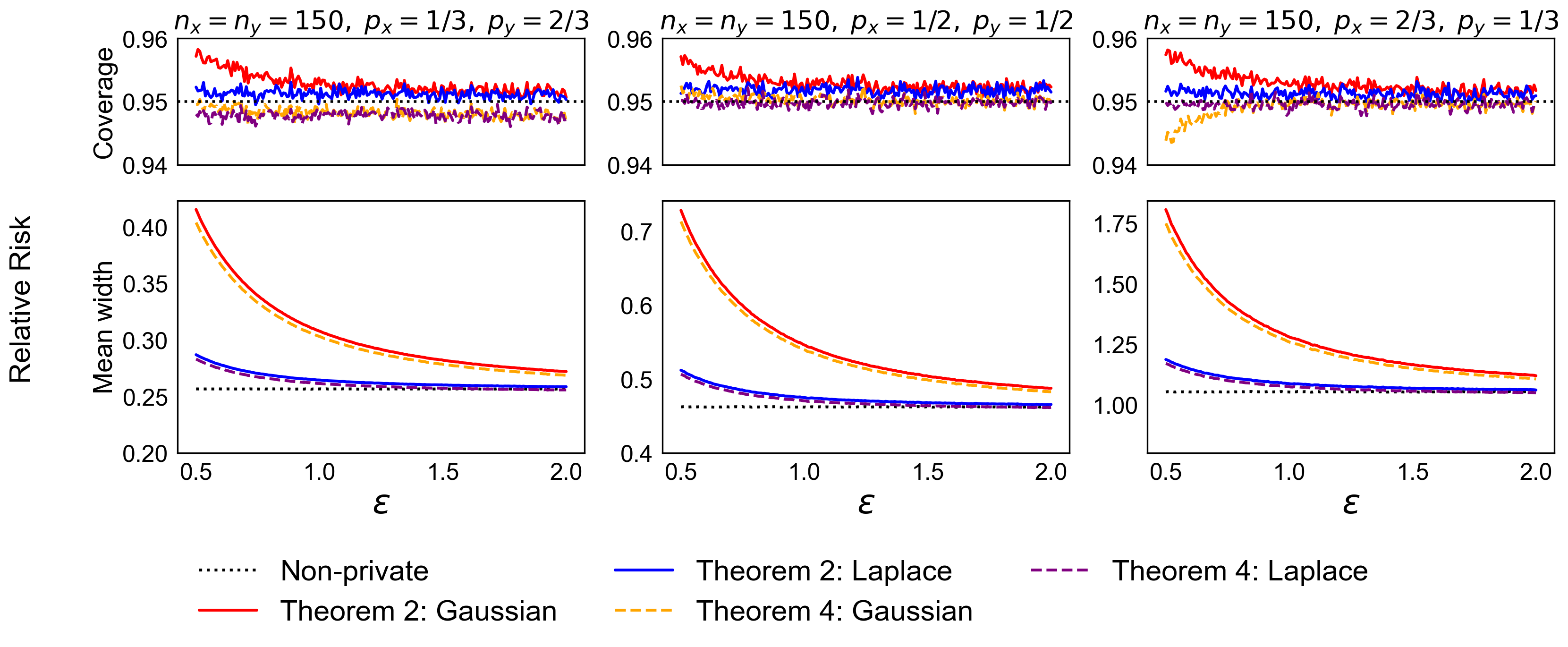}
\caption{\textbf{Relative Risk}: Mean width and coverage for CI construction. The dashed horizontal line marks the nominal coverage $1-\alpha$.}
\label{fig:CI_width_coverage_RR}
\end{subfigure}

\label{fig:CI_width_coverage_both}
\end{figure}

\vspace{-0.3cm}
\subsubsection*{Confidence interval width and coverage.} We evaluate the empirical coverage and average width of CIs for both the RR and the OR as functions of the privacy parameter $\varepsilon$ under the same setting as in the previous experiment.
For RR we compare two interval constructions: one based on Theorem~\ref{thm:ci_rr_gauss}, and the second on Theorem~\ref{thm:ci_rr_direct}. For OR, we use Theorem~\ref{thm:ci_or_gauss}. A noiseless baseline (setting the noise variances to zero) is included for reference.

Figures~\ref{fig:CI_width_coverage_OR} and \ref{fig:CI_width_coverage_RR} summarize the empirical coverage and mean width of the proposed CIs. Across all parameter configurations, the empirical coverage of the private intervals remains close to the nominal level $1-\alpha=0.95$. The behavior is stable even in asymmetric signal regimes, suggesting that the delta-method-based approximation remains accurate after noise addition.

As expected, the average width of the intervals decreases monotonically with $\varepsilon$. In the strong-privacy regime, perturbation variance dominates, leading to visibly wider intervals. As $\varepsilon$ increases, the widths approach those of the noiseless baseline, reflecting the fact that the additional variance from privacy becomes negligible relative to sampling variability. The Gaussian mechanism yields systematically wider intervals than the Laplace mechanism. 
This is expected, since for small $\delta$ the Gaussian noise scale is proportional to
$\sqrt{\log(1/\delta)}/\varepsilon$, whereas the Laplace scale is proportional to
$1/\varepsilon$. With $\delta=10^{-5}$, this constant factor is non-negligible at
moderate sample sizes.

Overall, the simulations confirm the theoretical predictions: even in these relatively small datasets, privacy-induced variability inflates interval width primarily when sampling variability is already substantial, while coverage remains well-controlled across privacy levels.

\subsubsection*{Real Data Application.}

\begin{table}[t]
\centering
\scriptsize
\setlength{\tabcolsep}{4pt}
\renewcommand{\arraystretch}{1.1}

\begin{subtable}[t]{0.32\textwidth}
\centering
\begin{tabular}{c|cc}
& \multicolumn{2}{c}{Myocardial Inf'}\\
\cline{2-3}
Group & Yes & No\\
\hline
Placebo & 189 & 10,845\\
Aspirin & 104 & 10,933\\
\end{tabular}
\caption{\footnotesize Table 2.1 in \cite{agresti2012categorical} (aggregated): Aspirin use vs.\ myocardial infarction.}
\label{tab:aspirin}
\end{subtable}
\hfill
\begin{subtable}[t]{0.32\textwidth}
\centering
\begin{tabular}{c|cc}
& \multicolumn{2}{c}{Breast cancer}\\
\cline{2-3}
Smoking & Case & Control\\
\hline
Ever smoked & 163 & 150\\
Never smoked & 124 & 161\\
\end{tabular}
\caption{\footnotesize Table 2 in \cite{rollison2008case}: tobacco smoking vs.\ breast cancer.}
\label{tab:smoking_breast_cancer}
\end{subtable}
\hfill
\begin{subtable}[t]{0.32\textwidth}
\centering
\begin{tabular}{c|cc}
& \multicolumn{2}{c}{Survival}\\
\cline{2-3}
Treatment & Yes & No\\
\hline
Streptomycin & 51 & 4\\
Control & 38 & 14\\
\end{tabular}
\caption{\footnotesize Table IX in \cite{mrc1948streptomycin}: MRC streptomycin RCT (tuberculosis).}
\label{tab:strep}
\end{subtable}

\caption{Three $2\times 2$ contingency tables used in the experiments.}
\label{tab:three_tables}
\end{table}
\begin{table}[t]
\centering
\scriptsize
\setlength{\tabcolsep}{3pt}
\renewcommand{\arraystretch}{1.12}

\begin{subtable}[t]{0.49\linewidth}
\centering
\caption{Relative risk (RR).}
\label{tab:rr_results_compact}
\begin{tabular}{@{}l c c c c c@{}}
\toprule
Panel & $\varepsilon$ & $\widehat{RR}$ & $\widetilde{RR}$ & Width & Width Priv \\
\midrule
\multirow{3}{*}{(a)}
& 0.5 & \multirow{3}{*}{0.550} & 0.551 & \multirow{3}{*}{0.264} & 0.298 \\
& 1   &                        & 0.550 &                        & 0.273 \\
& 2   &                        & 0.550 &                        & 0.266 \\
\addlinespace[2pt]

\multirow{3}{*}{(b)}
& 0.5 & \multirow{3}{*}{1.178} & 1.179 & \multirow{3}{*}{0.362} & 0.435 \\
& 1   &                        & 1.178 &                        & 0.381 \\
& 2   &                        & 1.178 &                        & 0.367 \\
\addlinespace[2pt]

\multirow{3}{*}{(c)}
& 0.5 & \multirow{3}{*}{1.269} & 1.282 & \multirow{3}{*}{0.461} & 140.928 \\
& 1   &                        & 1.271 &                        & 0.668 \\
& 2   &                        & 1.270 &                        & 0.519 \\
\bottomrule
\end{tabular}
\end{subtable}
\hfill
\begin{subtable}[t]{0.49\linewidth}
\centering
\caption{Odds ratio (OR).}
\label{tab:or_results_compact}
\begin{tabular}{@{}l c c c c c@{}}
\toprule
Panel & $\varepsilon$ & $\widehat{OR}$ & $\widetilde{OR}$ & Width & Width Priv \\
\midrule
\multirow{3}{*}{(a)}
& 0.5 & \multirow{3}{*}{0.546} & 0.546 & \multirow{3}{*}{0.265} & 0.299 \\
& 1   &                        & 0.546 &                        & 0.274 \\
& 2   &                        & 0.546 &                        & 0.268 \\
\addlinespace[2pt]

\multirow{3}{*}{(b)}
& 0.5 & \multirow{3}{*}{1.411} & 1.420 & \multirow{3}{*}{0.926} & 1.126 \\
& 1   &                        & 1.413 &                        & 0.979 \\
& 2   &                        & 1.412 &                        & 0.939 \\
\addlinespace[2pt]

\multirow{3}{*}{(c)}
& 0.5 & \multirow{3}{*}{4.697} & 9.086 & \multirow{3}{*}{13.977} & 61735.705 \\
& 1   &                        & 6.966 &                        & 1101.317 \\
& 2   &                        & 5.528 &                        & 44.206 \\
\bottomrule
\end{tabular}
\end{subtable}

\caption{Non-private and private point estimates and CI widths under Laplace-perturbed counts. For RR we use the log-scale CI of Theorem~\ref{thm:ci_rr_gauss}; for OR we use Theorem~\ref{thm:ci_or_gauss}.  Width denotes the noiseless CI width, and Width Priv denotes the average private CI width.}
\label{table:sum_real_data}
\end{table}

In this subsection, we illustrate the practical behavior of the proposed CIs using real $2\times 2$ contingency tables. We study both the RR and the OR, and compare their non-private and private releases to understand how DP noise affects statistical inference in practice. Since the RR intervals from Theorems ~\ref{thm:ci_rr_direct} and~\ref{thm:ci_rr_gauss} lead to very similar conclusions, we report only the interval of Theorem~\ref{thm:ci_rr_gauss} for simplicity.

We consider three representative $2\times2$ tables, displayed in Panels~(a)--(c) of Table~\ref{tab:three_tables}. These examples span three different regimes. Panel~(a) is a large-sample rare-event table: the total sample size is in the tens of thousands, while the event counts themselves are only in the hundreds. Panel~(b) is an intermediate-scale table, with all four cells in the hundreds and a moderate association between exposure and outcome. Panel~(c) is a much smaller table, with one cell in the single digits and a strong treatment effect. This progression from large counts, to moderate counts, to very small counts allows us to highlight how the magnitude of the cell counts governs the robustness of private inference. For the case-control table in Panel (b), the odds ratio is the epidemiologically natural estimand; we also report the RR-style plug-in ratio only as a numerical benchmark for the behavior of ratio statistics under privatization.

For each table, we release the two event counts as a two-dimensional count query and add independent Laplace noise with scale $b=1/\varepsilon$ to each coordinate. This vector release is $\varepsilon$-DP by the Laplace mechanism. We then construct CIs using Theorems~\ref{thm:ci_rr_gauss} and~\ref{thm:ci_or_gauss}, setting the perturbation variance to $\sigma^2 = 2b^2 = 2/\varepsilon^2$. For each dataset and each privacy level, we record the average value of the private estimator and the average width of the corresponding CI. These averages are computed over $N=10^5$ independent noise draws, so as to isolate the effect of the privacy mechanism itself. The results are summarized in Table~\ref{table:sum_real_data}.

Table~\ref{table:sum_real_data} shows that the impact of DP depends strongly on the scale of the underlying counts. In Panel~(a), the total sample size is very large, and even the event counts are in the hundreds, so the perturbation has only a minor effect. Both $\widetilde{RR}$ and $\widetilde{OR}$ are essentially identical to their non-private counterparts across all privacy levels, and the increase in interval width is modest. In Panel~(b), the effect of privacy is  visible, but remains well-controlled. The private point estimates are very close to the non-private ones for both RR and OR, and the CIs widen only moderately. Thus, when the cell counts are in the hundreds, Laplace perturbation introduces some additional uncertainty, but does not qualitatively distort the inferential conclusions. 

The most dramatic behavior appears in Panel~(c). In this case, the additive noise is no longer negligible relative to the observed counts, and the ratio estimators become highly unstable under strong privacy. For RR, the average private estimate increases only slightly, from $1.269$ to $1.282$ at $\varepsilon=0.5$, but the average private CI width explodes from $0.461$ to $140.928$. The effect is even more pronounced for OR: the average private CI width becomes enormous. This instability is exactly what one expects when noise can move a small count close to the boundary, causing ratios and log-ratios to vary dramatically.

\section{Conclusion}

We studied estimation and inference for relative risk and odds ratio when the underlying event counts are released under additive DP mechanisms. We showed how to construct plug-in estimators and variance-adjusted CIs that explicitly account for both sampling variability and privacy-induced noise. Empirically, the proposed intervals preserve coverage across a range of privacy levels, with widths close to their non-private counterparts in well-powered settings. Our variance decompositions further clarify when privacy affects inference: privacy has modest effects in high-count regimes, while sparse regimes remain fragile, both with and without privacy. These findings support simple plug-in inference from privatized counts as a practical tool, while also clarifying its limitations under sparse data.

% \section{Conclusion}

% We studied estimation and inference for ratio statistics—relative risk and odds ratio—when underlying counts are released under differential privacy via standard additive-noise mechanisms. We established asymptotic normality of the natural plug-in estimators computed from privatized counts (with optional clipping), and derived asymptotically valid confidence intervals that explicitly account for both sampling variability and privacy-induced perturbation. The resulting variance expressions decompose cleanly into a classical sampling term and a perturbation term, clarifying when privacy materially affects inference: perturbation is negligible when counts are large (high-power regimes) and can dominate only when cell counts are small, where non-private inference is already unstable.

% Simulations and real-data examples confirm these predictions, showing that for moderate privacy budgets and sufficiently large counts, private intervals closely match their non-private counterparts. 

\begin{credits}
\subsubsection{\ackname} This work was supported in part by a gift to the McCourt School of Public Policy and Georgetown University, Simons Foundation Collaboration 733792, Israel Science Foundation (ISF) grant 2861/20, Apple, ERC grant 101125913, and a grant from the Israeli Council of Higher Education. Views and opinions expressed are however those of the authors only and do not necessarily reflect those of the European Union or the European Research Council Executive Agency. Neither the European Union nor the granting authority can be held responsible for them.

\subsubsection{\discintname}
The authors have no competing interests to declare that are
relevant to the content of this article.
\end{credits}
%
% ---- Bibliography ----
%
% BibTeX users should specify bibliography style 'splncs04'.
% References will then be sorted and formatted in the correct style.
%
\clearpage
\bibliographystyle{splncs04}
\bibliography{references}

\newpage
\appendix
\begin{center}
\Large\textbf{Appendix}
\end{center}

\section{Proof of Theorem \ref{thm:ci_rr_gauss}}

\begin{proof}
We begin by showing that trimming the noisy counts has no asymptotic effect. Denote $\mu_x=p_x n_x$ and $\mu_y=p_y n_y$, and define
$
A_x := \{\widetilde X \notin [1,n_x-1]\},\qquad
A_y := \{\widetilde Y \notin [1,n_y-1]\}.
$
Since $p_x\in[\eta,1-\eta]$, we have $\min(\mu_x,n_x-\mu_x)\ge \eta n_x$. By the triangle inequality,
\[
\prob(A_x)
\le
\prob\!\big(|X-\mu_x|\ge \tfrac{\eta}{4}n_x\big)
+
\prob\!\big(|N_x|\ge \tfrac{\eta}{4}n_x\big).
\]
The first term tends to $0$ exponentially fast by Hoeffding's inequality for binomials. The second tends to $0$ because
$N_x\sim\mathcal N(0,\sigma_x^2)$ and $\sigma_x=o(n_x)$, implying
$\eta n_x/(4\sigma_x)\to\infty$.
Thus $\prob(A_x)\to0$, and similarly $\prob(A_y)\to0$. Consequently,
$
\prob(\widetilde X_p\neq \widetilde X)\to 0$,
 $\prob(\widetilde Y_p\neq \widetilde Y)\to 0,
$
and in particular
\begin{equation}\label{eq:proj_rr_clean}
\widetilde X_p/n_x\pto p_x,
\quad \widetilde Y_p/n_y \pto p_y.
\end{equation}
Next, we follow the construction of the delta method. Define
$
g(u,v)=\log\!\Big(\frac{u/n_x}{v/n_y}\Big).
$
The function $g$ is twice continuously differentiable in a neighborhood of $(\mu_x,\mu_y)$, with
$
\nabla g(\mu_x,\mu_y)=\left(\frac{1}{\mu_x},-\frac{1}{\mu_y}\right)$. Write
$$
\widetilde X-\mu_x=(X-\mu_x)+N_x, \quad
\widetilde Y-\mu_y=(Y-\mu_y)+N_y,
$$
where $(X,N_x)$ is independent of $(Y,N_y)$. Since $X\sim\mathrm{Bin}(n_x,p_x)$,
$
\frac{X-\mu_x}{\sqrt{\Var(X)}} \dto \mathcal N(0,1)$,
$\Var(X)=n_xp_x(1-p_x),
$
and analogously for $Y$. To identify the asymptotic law of $\widetilde X-\mu_x$, write
\[
\frac{\widetilde X-\mu_x}{\sqrt{\Var(X)+\sigma_x^2}}
=
a_{n,x}\frac{X-\mu_x}{\sqrt{\Var(X)}}
+
b_{n,x}\frac{N_x}{\sigma_x},
\]
where
\[
a_{n,x}
=
\sqrt{\frac{\Var(X)}{\Var(X)+\sigma_x^2}},
\qquad
b_{n,x}
=
\sqrt{\frac{\sigma_x^2}{\Var(X)+\sigma_x^2}}.
\]
Let
\[
U_{n,x}=\frac{X-\mu_x}{\sqrt{\Var(X)}},
\qquad
G_x=\frac{N_x}{\sigma_x}\sim \mathcal N(0,1).
\]
Then $U_{n,x}\dto \mathcal N(0,1)$ by the classical CLT, and $G_x$ is independent of $U_{n,x}$. Hence, for each fixed $t\in \R$, the characteristic function of $a_{n,x}U_{n,x}+b_{n,x}G_x$
is
\[
\varphi_n(t)
=
\E e^{it a_{n,x}U_{n,x}}
\cdot
\E e^{it b_{n,x}G_x}
=
\varphi_{U_{n,x}}(a_{n,x}t)
\exp\!\left(-\frac{b_{n,x}^2t^2}{2}\right).
\]
Since $U_{n,x}\dto \mathcal N(0,1)$, its characteristic functions converge to
$s\mapsto \exp(-s^2/2)$, uniformly for $s$ in compact sets; see, e.g.,
Lévy's continuity theorem \cite[Sec.~26]{billingsley1995probability}.
Because $a_{n,x}\in[0,1]$, this gives
$
\varphi_{U_{n,x}}(a_{n,x}t)
=
\exp\!\left(-\frac{a_{n,x}^2t^2}{2}\right)+o(1).
$
Therefore,
\[
\varphi_n(t)
=
\exp\!\left(-\frac{(a_{n,x}^2+b_{n,x}^2)t^2}{2}\right)+o(1)
=
\exp\!\left(-\frac{t^2}{2}\right)+o(1),
\]
where we used $a_{n,x}^2+b_{n,x}^2=1$. By Lévy's continuity theorem,
\[
\frac{\widetilde X-\mu_x}{\sqrt{\Var(X)+\sigma_x^2}}
=
a_{n,x}U_{n,x}+b_{n,x}G_x
\dto
\mathcal N(0,1).
\]
The same argument applies to $\widetilde Y$.

Let $s_x^2=\Var(X)+\sigma_x^2$ and $s_y^2=\Var(Y)+\sigma_y^2$. 
Define
$$
V_{n_x,n_y}
=
\frac{s_x^2}{\mu_x^2}
+
\frac{s_y^2}{\mu_y^2}
=
\left(\frac{1}{n_xp_x}-\frac{1}{n_x}\right)
+
\left(\frac{1}{n_yp_y}-\frac{1}{n_y}\right)
+
\frac{\sigma_x^2}{n_x^2p_x^2}
+
\frac{\sigma_y^2}{n_y^2p_y^2}.
$$
Since
$s_x/n_x\to0$ and $s_y/n_y\to0$, a Taylor expansion around $(\mu_x,\mu_y)$ gives
\[
\widetilde\ell-\ell^*
=
\nabla g(\mu_x,\mu_y)^\top
\begin{pmatrix}
\widetilde X-\mu_x\\
\widetilde Y-\mu_y
\end{pmatrix}
+
O_p\!\left(\frac{s_x^2}{n_x^2}+\frac{s_y^2}{n_y^2}\right).
\]
By independence of the two counts and their perturbations, the leading linear term is asymptotically normal with variance $V_{n_x,n_y}$.
Moreover,
$$
\frac{s_x^2}{n_x^2}+\frac{s_y^2}{n_y^2}
=
o\!\left(
\sqrt{
\frac{s_x^2}{n_x^2}
+
\frac{s_y^2}{n_y^2}
}
\right),
$$
because $s_x/n_x\to0$ and $s_y/n_y\to0$. Hence the Taylor remainder is
$o_p(\sqrt{V_{n_x,n_y}})$.

Finally, by \eqref{eq:proj_rr_clean} and continuity of $t\mapsto 1/t$ and $t\mapsto 1/t^2$ on $(0,\infty)$,
$
\widehat{\mathrm{se}}^2/V_{n_x,n_y}\pto 1.
$
Slutsky's theorem therefore implies
$
\frac{\widetilde\ell-\ell^*}{\widehat{\mathrm{se}}}
\dto
\mathcal N(0,1),
$
which yields
$$
\prob\!\Big(\ell^*\in[\widetilde\ell \pm z_{1-\alpha/2}\widehat{\mathrm{se}}]\Big)\to 1-\alpha.
$$
Exponentiating the endpoints preserves coverage for $\mathrm{RR}=p_x/p_y$.
\end{proof}

\section{Proof of Theorem \ref{thm:ci_or_gauss}}

\begin{proof}
The trimming argument is identical to the beginning of the proof of Theorem~\ref{thm:ci_rr_gauss}. In particular, clipping has no asymptotic effect, and
\begin{equation}\label{eq:proj_or_clean}
\widetilde X/n_x\pto p_x,
\qquad
\widetilde Y/n_y\pto p_y.
\end{equation}
Moreover, by the same argument as in that proof,
$$
\frac{\widetilde X-\mu_x}{\sqrt{\Var(X)+\sigma_x^2}} \dto \mathcal N(0,1),
\qquad
\frac{\widetilde Y-\mu_y}{\sqrt{\Var(Y)+\sigma_y^2}} \dto \mathcal N(0,1),
$$
where $\mu_x=p_xn_x$, $\mu_y=p_yn_y$.

We now apply the delta method with
$
g(u,v)=\log\!\Big(\frac{u}{n_x-u}\Big)-\log\!\Big(\frac{v}{n_y-v}\Big).
$
The function $g$ is twice continuously differentiable in a neighborhood of
$(\mu_x,\mu_y)$, with
\[
\nabla g(\mu_x,\mu_y)
=
\left(
\frac{1}{\mu_x}+\frac{1}{n_x-\mu_x},
-\Big(\frac{1}{\mu_y}+\frac{1}{n_y-\mu_y}\Big)
\right).
\]
Hence
$$
\widetilde\ell-\ell^*
=
\nabla g(\mu_x,\mu_y)^\top
\begin{pmatrix}
\widetilde X-\mu_x\\
\widetilde Y-\mu_y
\end{pmatrix}
+
R_{n_x,n_y},
$$
Denote
$$
s_x^2 \coloneqq \Var(X)+\sigma_x^2,
\qquad
s_y^2 \coloneqq \Var(Y)+\sigma_y^2.
$$
The leading linear term has variance
$$
V_{n_x,n_y}
=
\left(\frac{1}{\mu_x}+\frac{1}{n_x-\mu_x}\right)^2 s_x^2
+
\left(\frac{1}{\mu_y}+\frac{1}{n_y-\mu_y}\right)^2 s_y^2,
$$
and the remainder satisfies
$
R_{n_x,n_y}=o_p\left(\sqrt{V_{n_x,n_y}}\right).
$

Indeed, on an event whose probability tends to one, $\widetilde X$ and
$\widetilde Y$ remain in fixed interior fractions of their ranges, and the second
derivatives of $g$ are of order $O(n_x^{-2})$ and $O(n_y^{-2})$.
Moreover, since $s_x/n_x\to0$ and $s_y/n_y\to0$, the displayed remainder is
$o_p(\sqrt{V_{n_x,n_y}})$. By independence of the two counts and their perturbations, the leading linear term is asymptotically normal with variance
\begin{align*}
V_{n_x,n_y}
&=
\Big(\frac{1}{\mu_x}+\frac{1}{n_x-\mu_x}\Big)^2\big(\Var(X)+\sigma_x^2\big)
+
\Big(\frac{1}{\mu_y}+\frac{1}{n_y-\mu_y}\Big)^2\big(\Var(Y)+\sigma_y^2\big)\nonumber\\
&=
\frac{1}{n_xp_x(1-p_x)}+\frac{1}{n_yp_y(1-p_y)}
+
\frac{\sigma_x^2}{n_x^2}\Big(\frac{1}{p_x}+\frac{1}{1-p_x}\Big)^2
+
\frac{\sigma_y^2}{n_y^2}\Big(\frac{1}{p_y}+\frac{1}{1-p_y}\Big)^2.
\label{eq:V_or_clean}
\end{align*}
Therefore
$
\frac{\widetilde\ell-\ell^*}{\sqrt{V_{n_x,n_y}}}
\dto
\mathcal N(0,1)
$, and by \eqref{eq:proj_or_clean} and continuity of $t\mapsto 1/t$ and $t\mapsto 1/(1-t)$ on $(0,1)$,
$
\widehat{\mathrm{se}}^2/V_{n_x,n_y}\pto 1.
$
Slutsky's theorem then yields
$
\frac{\widetilde\ell-\ell^*}{\widehat{\mathrm{se}}}\dto \mathcal N(0,1),
$
and therefore
$$
\prob\!\Big(\ell^*\in[\widetilde\ell \pm z_{1-\alpha/2}\widehat{\mathrm{se}}]\Big)\to 1-\alpha.
$$
Exponentiating the endpoints preserves coverage for $\mathrm{OR}$.
\end{proof}

\end{document}